\documentclass[
 aip,
 cha,
 amsmath,amssymb,
reprint,
]{revtex4-1}

\usepackage{graphicx}
\usepackage{dcolumn}
\usepackage{bm}
\usepackage[utf8]{inputenc}
\usepackage[T1]{fontenc}
\usepackage{mathptmx}
\usepackage{etoolbox}
\usepackage{hyperref}
\usepackage{overpic}
\usepackage{xcolor}  
\usepackage{epstopdf}

\makeatletter
 \def\@email#1#2{
 \endgroup
 \patchcmd{\titleblock@produce}
  {\frontmatter@RRAPformat}
  {\frontmatter@RRAPformat{\produce@RRAP{*#1\href{mailto:#2}{#2}}}\frontmatter@RRAPformat}
 {}{}
}
\makeatother
\begin{document}

\preprint{AIP/123-QED}

\title{Evolutionary dynamics in state-feedback public goods games with peer punishment}

\author{Qiushuang Wang}
\author{Xiaojie Chen}
\email{xiaojiechen@uestc.edu.cn}
\altaffiliation[Corresponding author]{}
\affiliation{
School of Mathematical Sciences, University of Electronic Science and Technology of China, Chengdu 611731, China}
\author{Attila Szolnoki}
\affiliation{Institute of Technical Physics and Materials Science, Centre for Energy Research, P.O. Box 49, H-1525 Budapest, Hungary}

\date{\today}

\begin{abstract}
Public goods game serves as a valuable paradigm for studying the challenges of collective cooperation in human and natural societies. Peer punishment is often considered as an effective incentive for promoting cooperation in such contexts. However, previous related studies have mostly ignored the positive feedback effect of collective contributions on individual payoffs. In this work, we explore global and local state-feedback, where the multiplication factor is  positively correlated with the frequency of contributors in the entire population or within the game group, respectively. By using replicator dynamics in an infinite well-mixed population we reveal that state-based feedback plays a crucial role in alleviating the cooperative dilemma by enhancing and sustaining cooperation compared to the feedback-free case. Moreover, when the feedback strength is sufficiently strong or the baseline multiplication factor is sufficiently high, the system with local state-feedback provides full cooperation, hence supporting the ``think globally, act locally'' principle. Besides, we show that the second-order free-rider problem can be partially mitigated under certain conditions when the state-feedback is employed. Importantly, these results remain robust with respect to variations in punishment cost and fine.
\end{abstract}

\maketitle

\begin{quotation}
The emergence and maintenance of cooperation among selfish individuals remains an open enigma. Evolutionary game theory provides an important and effective theoretical framework for studying this puzzle. Particularly, the conflicting interactions of multiple individuals in the real world can be modeled using the public goods game. Besides, peer punishment is an important incentive to promote cooperation in such conflicts. However, existing research often overlooks a critical factor: the feedback effect of collective contributions on individual payoffs. To address this gap, we introduce global and local state-feedback by considering that the multiplication factor of collective efforts depends on the local and global proportions of contributors, respectively. The consequence of this feedback is studied in a well-mixed system where participants play a public goods game with punishment. Our theoretical analysis demonstrates that, regardless of the punishment cost or fine, the synergy between state-based feedback, particularly local state-feedback and punishment significantly enhances the evolution of cooperation. Moreover, state-based feedback can help mitigate the second-order free-rider problem under certain conditions.
\end{quotation}

\section{\label{sec1}INTRODUCTION}

Cooperation is common both in human and natural societies, although it incurs individual cost. Understanding how cooperation emerges and is maintained among selfish individuals has long been a pivotal topic~\cite{Pennisi_Science_2005}.
Evolutionary game theory provides a significant and effective theoretical framework for studying the evolution of cooperation~\cite{Smith_CambridgeUP_1982,Hofbauer_CambridgeUP_1998,Nowak_HarvardUP_2006}.
Therein, the public goods game is the most widely used paradigm to characterize the interaction process between multiple individuals~\cite{Roth_PrinvetonUP_1995,Hauert_JTB_2006,Jesus_Chaos_2011,Szolnoki_JTB_2013,Han_JRSI_2015,Liu_Chaos_2018,Chen_NJP_2015,Wang_AMC_2022,Duh_PRE_2020,Santos_Nature_2008,Han_AAMAS_2017}.

In a classical public goods game, each individual decides whether to contribute to a common pool or not. Accordingly, cooperators invest at the expense of their individual cost, and the total investments are enhanced by a multiplication factor. The result is equally divided among participants, regardless of their contributions. Therefore, defectors, who choose not to contribute, receive a higher benefit than cooperators. Under natural selection, individuals tend to defect, which results in the tragedy of the commons state~\cite{Hardin_68}. To address this dilemma, various extensions of the classical public goods game have been proposed, including the optional public goods game~\cite{Hauert_Science_2002}, the nonlinear public goods game~\cite{Milinski_PNAS_2008,Santos_PNAS_2011,Chica_SR_2021,Wang_PhysicaD_2019}, and others. One particularly significant class of nonlinear public goods games is where the multiplication factor depends on the environmental state~\cite{Shao_EPL_2019}, the group size \cite{Lee_ChaosSF_2023}, or the frequency of cooperators \cite{Shi_PhysicaA_2012,Ma_AMC_2023,Hauert_JTB_2006,Chen_PloSone_2012,Weng_PLA_2021,Hauert_JRSI_2024}.

In parallel, several other mechanisms have been considered to explain the evolution of cooperation~\cite{Santos_Nature_2008,Allen_elife_2013,Allen_Nature_2017,Wu_SR_2016,Hauert_JTB_2016,Sasaki_PRSB_2013,Santos_PRSB_2015,Sasaki_BL_2014,Sasaki_CRR_2015,Duong_PRSA_2021,WJ_NonlinearDyn_2024,Li_AMC_2024,Li_CSF_2024}. For example, the usage of incentives, like punishing defectors have been proved as an effective tool to promote cooperation~\cite{Nakao_BP_2012,Fehr_AER_2000,Gurerk_Science_2006,Huang_SR_2018,Rand_NC_2011,Sasaki_PRSB_2013,Traulsen_Science_2007,Hauert_DCDSSB_2004,Chen_JRSI_2015,Hauert_JTB_2002}. In the public goods game with peer punishment, individuals can choose to act as prosocial punishers, who contribute not only to the common pool but also penalize defectors. Accordingly, the introduction of peer punishment can effectively promote cooperation, which was verified by behavioral experiments or theoretical analyses~\cite{Taylor_Evolution_2007,Fehr_Nature_2002,Boyd_ES_1992,Gachter_Science_2008,Boyd_PNAS_2003,Fowler_PNAS_2005}. However, since peer punishment is costly, cooperators, who also enjoy the benefit of punishment, do better than punishers hence creating a second-order free-rider problem where cooperators become second-order free-riders. This restores the original dilemma therefore understanding how punishment can work attracted significant scientific interest~\cite{Traulsen_Science_2007,Chen_NJP_2014,Perry_JSP_2018,Helbing_PloSCB_2010,Wang_AMC_2018}.

Nevertheless, these works have ignored the fact that a greater number of contributors can provide the significantly higher return of the public goods~\cite{Bejan_JAP_2017,Mauler_AE_2021,Straus_JAE_2022}. Specifically, as the size of the global proportion of contributors in the entire population or  the local proportion of contributors within the game group increases, the overall public benefit grows, which in turn raises the marginal per capita payoff for each additional contributor (i.e., the multiplication factor), creating a positive feedback loop. We refer to the feedback arising from global or local collective contributions on individual payoffs as global or local state-feedback, respectively. Such a state-based feedback is ubiquitous both in human societies and in the natural world, such as in economies of scale \cite{Bejan_JAP_2017,Mauler_AE_2021}. However, it remains unclear whether the synergy of state-based feedback and peer punishment can solve the cooperation dilemma and alleviate the above-mentioned second-order free-rider problem.

Motivated by this shortage of the theory, we respectively consider that the multiplication factor varies linearly with the frequency of contributors in the whole population or with the fraction of contributors in the local group. We then use the replicator equation approach to study the evolutionary dynamics in the state-feedback public goods game with peer punishment in infinite well-mixed populations. Through  stability analyses, we demonstrate that systems with state-based feedback exhibit more favorable dynamics for the emergence and maintenance of cooperation. This effect is especially salient when local state-feedback is on stage. Crucially, when  feedback strength is sufficiently strong or the baseline multiplication factor is large enough, the system can evolve toward and stabilize in a full-cooperation state. Additionally, we show that the introduction of state-based feedback, whether it is global or local, alleviates the second-order free-rider problem in the system under certain conditions. We also stress that these results remain robust even with variations in punishment cost and fine.

\section{\label{sec2}MODEL AND METHODS}
\subsection{\label{sec2_1}Modeling state-based feedbacks in public goods game with peer punishment}

We consider an infinite well-mixed population, where each individual can choose to be a cooperator ($C$, who contributes but does not punish), a defector ($D$, who does not contribute), or a punisher ($P$, who contributes and simultaneously punishes all defectors in the group).

In each step, $n$ $(n\ge 2)$ individuals are randomly chosen from the population to form a group and play the public goods game. Assume that  among these $n$ players, there are $n_C$ cooperators, $n_D$ defectors, and $n_P$ punishers. Each contributor (cooperator or punisher) contributes $c$ to the public pool, while a defector does not. The total contributions are multiplied by a multiplication factor $r$ and the result is shared equally among all $n$ participants. Besides, each punisher imposes a $\beta$ fine on each defector for a $0< \gamma < \beta$ price. This extra cost of punishment is borne by the punisher.

We then consider the global ($G$) or local ($L$) state-feedback on the multiplication factor. For the global state-feedback, we assume that
\begin{equation}\label{r_1}
	r_{G} = r_{0} + k(x + z)\,,
\end{equation}
where $x$ and $z$ represent the fraction of cooperators and punishers in the whole population, respectively.

For the local state-feedback, we assume that
\begin{equation}\label{r_2}
	r_{L} = r_{0} + k\frac{n_C+n_P}{n}\,.
\end{equation}
Here, $k>0$ serves as the feedback strength, and $r_{0}>1$ denotes the baseline multiplication factor, reflecting the minimum level of multiplication in the public goods game. The modified multiplication factors presented in Eqs.~\eqref{r_1} and \eqref{r_2} clearly capture the linear dependence of the feedback return of contributions to the public pool on the frequency of contributors, highlighting positive feedback based on either global or local states, respectively. In the rest of this work we maintain the condition $1<r_0\le r_{j}\le r_{0}+k<n$ ($j\in\left\lbrace G,L\right\rbrace $), hence ensuring the presence of a social dilemma in the absence of punishment~\cite{Chen_JRSI_2015}.

Consequently, the payoffs of cooperators, defectors, and punishers in the interaction group are given as
\begin{equation*}
	\Pi_{C}(n_C,n_D,n_P)=\frac{r_{j}(n_C+n_P)}{n}c-c,
\end{equation*}
\begin{equation*}
	\Pi_{D}(n_C,n_D,n_P)=\frac{r_{j}(n_C+n_P)}{n}c-\beta n_P,
\end{equation*}
and
\begin{equation*}
	\Pi_{P}(n_C,n_D,n_P)=\frac{r_{j}(n_C+n_P)}{n}c-c-\gamma n_D,
\end{equation*}respectively, where $j\in\left\lbrace G,L\right\rbrace $.

\subsection{\label{sec2_2}Replicator dynamics}

We consider that the player with the lower payoff adopts the strategy of the more successful player, with a probability proportional to the payoff difference. Thus the evolution of strategies in infinite well-mixed populations can be described by the replicator equations~\cite{Hofbauer_CambridgeUP_1998}:
\begin{equation}
	\left\{
	\begin{aligned}
		\dot{x}&=x(P_{C}-\overline{P}), \\
		\dot{y}&=y(P_{D}-\overline{P}), \\
		\dot{z}&=z(P_{P}-\overline{P}),
	\end{aligned}
	\right.
\end{equation}
where $y\triangleq 1-x-z$ is the fraction of defectors in the population, $P_i$  ($i\in\left\lbrace C,D,P\right\rbrace  $) is the expected payoff for individuals who adopt strategy $i$, and $\overline{P}$ denotes the average payoff in the whole population, given by $\overline{P}=xP_{C}+yP_{D}+zP_{P}$.

Specially, for a focal individual, let $m$ and $s$ denote the number of cooperators and defectors among the other $n-1$ co-players in the game group, respectively. Hence, the number of punishers among the other $n-1$ co-players in the game group are $n-1-m-s$. Therefore, the expected payoffs for each strategy are
\begin{equation}\label{average payoff1}
	\begin{split}
		P_{C}
		=&
		\sum_{m=0}^{n-1}\sum_{s=0}^{n-1-m}\binom{n-1}{m}\binom{n-1-m}{s}x^{m}y^{s}z^{n-1-m-s}\\
		&\cdot  \vphantom{\frac{2}{3}} \Pi_{C}(m+1,s,n-1-m-s)\,,
	\end{split}
\end{equation}
\begin{equation}\label{average payoff2}
	\begin{split}
		P_{D}
		=&
		\sum_{m=0}^{n-1}\sum_{s=0}^{n-1-m}\binom{n-1}{m}\binom{n-1-m}{s}x^{m}y^{s}z^{n-1-m-s}\\
		&\cdot  \vphantom{\frac{2}{3}} \Pi_{D}(m,s+1,n-1-m-s)\,,
	\end{split}
\end{equation}
and
\begin{equation}\label{average payoff3}
	\begin{split}
		P_{P}
		=&
		\sum_{m=0}^{n-1}\sum_{s=0}^{n-1-m}\binom{n-1}{m}\binom{n-1-m}{s}x^{m}y^{s}z^{n-1-m-s}\\
		&\cdot \vphantom{\frac{2}{3}} \Pi_{P}(m,s,n-s-m)\,.
	\end{split}
\end{equation}
	
Accordingly, the replicator dynamics with global state-feedback are described by the following differential equation system:
\begin{equation}\label{sys_1}
	\left\{
	\begin{aligned}
		\dot{x}=&x(1-x-z)\left[z (\beta+\gamma)(n-1)+\frac{r_{0}+k(x+z)}{n}c -c \right], \\
		\dot{z}=&z(1-x-z)\left[ z(\beta+\gamma)(n-1)+\frac{r_{0}+k(x+z)}{n}c \right.  \\
		& \left.-c -\gamma(n-1)\vphantom{\sum_{a}}\right].
	\end{aligned}
	\right.
\end{equation}
For local state-feedback, the replicator equations are given by
\begin{equation}\label{sys_2}
	\left\{
	\begin{aligned}
		\dot{x}=&x(1-x-z)\left[z (\beta+\gamma)(n-1)+\frac{2ck}{n^{2}}(x+z)\left(n -1\right)\right.  \\
		& \left.+\frac{r_{0}}{n}c+\frac{k}{n^{2}}c-c \right], \\
		\dot{z}=&z(1-x-z)\left[z (\beta+\gamma)(n-1)+\frac{2ck}{n^{2}}(x+z)\left(n -1\right)\right.  \\
		& \left.+\frac{r_{0}}{n}c+ \frac{k}{n^{2}}c-c -\gamma(n-1)\right].
	\end{aligned}
	\right.
\end{equation}
Regardless of whether it is a local state or global state-feedback, the system reverts to the system without state-based feedback when $k=0$~\cite{Sasaki_PRSB_2013,Hauert_DCDSSB_2004}, as characterized by Eq.~\eqref{sys_0} in Appendix~\ref{suppA}.

In our model, linear positive feedback based on local or global states enhances the return for  public goods, thereby increasing the payoffs of individuals participating in the public goods game. As a result, individuals, regardless of their strategy, achieve higher payoffs with state-based feedback compared to the feedback-free case. However, it remains unclear whether the punishment mechanism can still effectively promote cooperation once state-based feedback is introduced. In Section \ref{sec3}, we will present the evolutionary dynamics of systems with and without state-based feedback and make comparisons. More importantly, we will focus on  whether global or local state-feedback more effectively fosters the evolution of cooperative behavior and relieves the issue of second-order free riding.

\section{\label{sec3}RESULTS}
\begin{figure*}
	\includegraphics{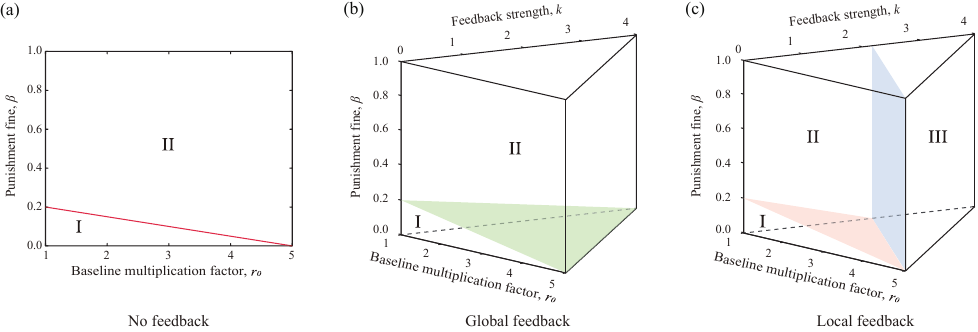}
	\caption{\label{fig1}Schematic diagrams illustrating the evolutionary outcomes of systems with and without state-based feedback. In region~I, the system exhibits a single asymptotically stable state where regardless of the initial state the system will eventually converge to a full defection state.	For region~II and III, cooperation can be maintained. In region~II, the system possesses the continuous set of stable and unstable equilibria, all representing that every individual will invest in the public pool. However, the state where all individuals defect remains absorbing. In region~III, all states of full contribution are absorbing, indicating that the system achieves a higher level of cooperation, although the state of full defection still remains absorbing. The red, green, pink, and blue lines or surfaces, representing the boundaries between different regions, correspond to our analytical conditions, given by $z^*_{N}=1$, $ z^*_{G}=1$, $ z^*_{L}=1$, and $ z^*_{L}=0$. Parameters: $n=5$, $ c=1$, and $\gamma=0.01$.}
\end{figure*}
\begin{figure*}
	\includegraphics{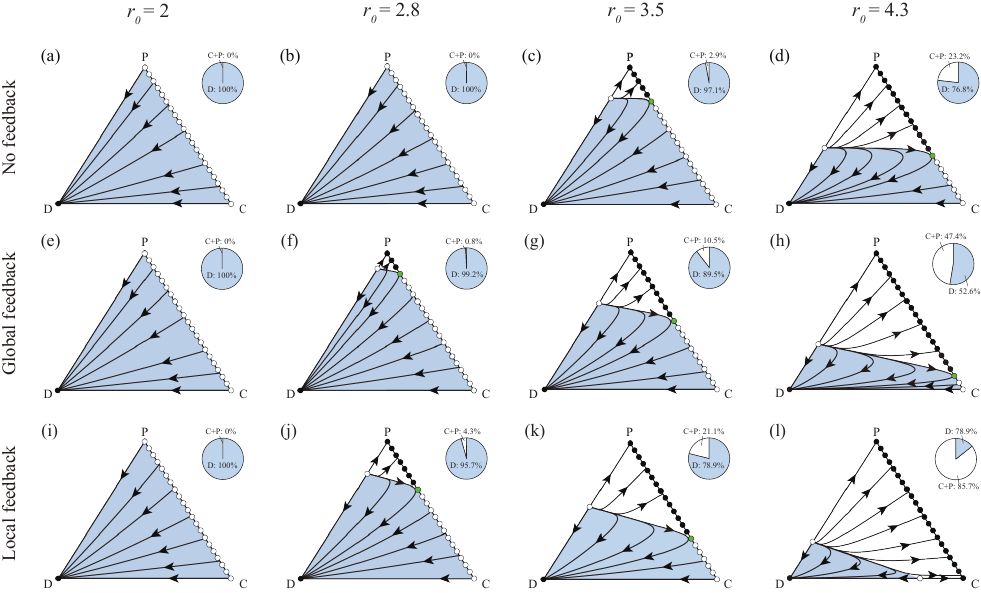}
	\caption{\label{fig2}Evolutionary dynamics on the phase planes of the system under varying $r_0$. The solid and open black circles depict stable and unstable equilibria, respectively. The green dots indicate the critical point $(x,z)=(1-z^*_{l}, z^*_{l})$, where $l\in\left\lbrace N, G, L \right\rbrace $. In each phase plane, the blue region represents the attraction basin of the stable full-defection equilibrium (corresponding to the lower-left vertex of each simplex $S_{3}$). Consequently, the remaining white area represents the attraction basin of the stable full-contribution equilibria. Additionally, the precise proportions of these two attraction domains are given in each corresponding pie chart, where `D' represents a defector-dominated state and `C+P' represents a contributor-dominated state. Parameters in panels (a)-(d): $n=5$, $\beta=0.1$, $ c=1$, and $\gamma=0.01$. Parameters in panels (e)-(l): $n=5$, $\beta=0.1$, $ c=1$, $\gamma=0.01$, and $k=0.5$.}
\end{figure*}
By analyzing the local stability of the equilibria of Eqs.~\eqref{sys_1} and \eqref{sys_2}, and comparing them to the feedback-free system Eq.~\eqref{sys_0}~\cite{Sasaki_PRSB_2013,Hauert_DCDSSB_2004}, we present schematic diagrams that depict the evolutionary outcomes of systems with and without state-based feedback in Fig.~\ref{fig1}.
As shown in Fig.~\ref{fig1}(a), for the classical public goods game with punishment (i.e., without state-feedback), the system can exhibit two distinct dynamic behaviors depending on the values of $r_{0}$ and $\beta$. When $r_0$ and $\beta$ are small, there is a unique stable equilibrium at $(0,0)$, indicating that, regardless of the initial state, the system will evolve toward a state dominated by defectors (corresponding to the dynamics of region~I). For a larger $r_{0}$ or $\beta$, the equilibrium $(0,0)$ and the continuous set of equilibria $(x,z)$ (where $x+z=1$  and $z > z^*_{N}$) are stable, where
\begin{equation}\nonumber
	z^{*}_{N}=\frac{ n-r_{0}}{\beta n(n-1)}c,
\end{equation} as described in Appendix~\ref{suppA}.
This implies that the system will ultimately evolve toward either a state of all defectors or a state of all contributors, with at least $z^*_N$ punishers (corresponding to the dynamics of region~II). When global state-feedback is introduced, the system's evolutionary outcomes remain unchanged, as shown in Fig.~\ref{fig1}(b). However,  in region~II of Fig.~\ref{fig1}(b), a minimum of $z^*_{G}$ punishers is required to achieve a stable full-contribution state (a more detailed description is provided in Appendix~\ref{suppB}), where
\begin{equation}\nonumber
	z^{*}_{G}=\frac{n-r_{0}-k}{\beta n(n-1)}c.
\end{equation}
When local state-feedback is considered, as presented in Fig~\ref{fig1}(c), the evolutionary dynamics shown in regions~I and II can still be achieved. Besides, in region~II of Fig.~\ref{fig1}(c), the minimum number of punishers required to reach the stable full-contribution state is given by
\begin{equation}\nonumber
	z^{*}_{L}=\frac{(n^{2}-nr_{0}-k)-2k(n-1)}{\beta n^{2}(n-1)}c,
\end{equation}
as explained in Appendix~\ref{suppC}. Unexpectedly, local state-feedback can lead to more favorable evolutionary dynamics when $r_0$ or $k$ is large. In region~III of Fig.~\ref{fig1}(c), the point $(0,0)$ remains a stable equilibrium, while all points along the continuum $(x, z)$ (where $x + z = 1$, representing varying proportions of cooperators and punishers) are also stable equilibria. The system  will eventually evolve to either the state of all defectors or all contributors. Notably,  the full-cooperation state is locally stable, meaning cooperation can dominate and persist. Additionally, since the boundaries of different regions are determined by $z^*_l$ (where $l\in\left\lbrace N,G,L \right\rbrace$) and $z_l^*$ is independent of the parameter $\gamma$, the dynamics of the system in each region will remain unchanged as $\gamma$ varies. Therefore, we demonstrate that  the system with local state-feedback displays richer dynamic behaviors than systems with no feedback or global feedback.  Moreover, it can achieve a full-cooperation state under certain conditions (a more detailed analysis is provided in the Appendix \ref{suppA}--\ref{suppC}).
\begin{figure*}
	\includegraphics{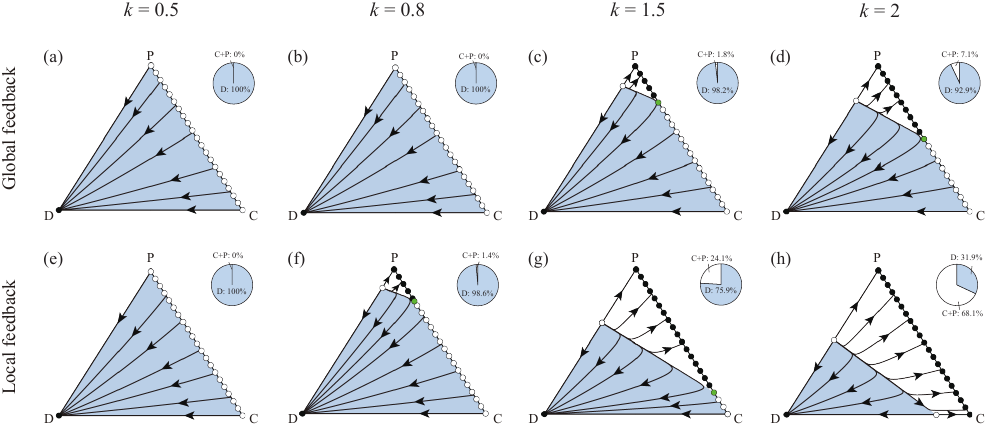}
	\caption{\label{fig3}Evolutionary dynamics on the phase planes of the systems for different feedback strength.
		The solid and open black circles represent stable and unstable equilibria, respectively. The green dots indicate the critical point $(x,z)=(1-z^*_{l},z^*_{l})$, where $l\in\left\lbrace N, G, L \right\rbrace $. The blue region denotes the attraction basin of the stable equilibrium where all individuals defect. The remaining white area represents the attraction basin of the stable equilibria where all individuals contribute.
		Parameters: $n=5$, $\beta=0.1$, $ c=1$, $\gamma=0.01$, and $r_{0}=2$.}
\end{figure*}

We next present a detailed analysis of dynamics for these three cases,
and examine how the baseline multiplication factor $r_0$ affects the evolutionary dynamics of the system, both with and without state-based feedback, as shown in Fig.~\ref{fig2}.
Fig.~\ref{fig2} presents phase diagrams within the simplex $ S_3$  for different values of $r_0$. The blue and white  areas represent the attractive domains of stable full-contribution and full-defection states, respectively. And, the boundaries between the two attractive domains for the three red cases, such as no feedback, global feedback, and local feedback, are determined by the real roots $(x, z)\in [0,1]\times[0,1]$ (with $x+z\le1$) of the functions $f_{l}(x,z)$ (where $l\in\left\lbrace N,G,L\right\rbrace $), which are detailed in the Appendix \ref{suppA}--\ref{suppC}. We observe that as $r_0$  increases, no matter whether or not the state-based feedback on the multiplication factor is considered, continuously stable full-contribution equilibria emerge, with an increasingly larger attractive domain. Concretely, in the absence of feedback (see the first row of Fig.~\ref{fig2}), for relatively small values of $r_0$, the system exhibits a  single stable  full defection state. However, when $r_0$ is large (e.g., for $r_0=3.5$ and $r_0=4.3$), stable full-contribution equilibria emerge, promoting the evolution of cooperation. Nevertheless, the point $(0,0)$ remains a stable equilibrium for any value of $r_0$, indicating that the system may still fall into the tragedy of the commons state. Fortunately, state-based feedback can reduce the likelihood of this outcome (see the second and third rows of Fig.~\ref{fig2}). When the state-based feedback is considered, as $r_{0}$ increases, the stable full-contribution state appears earlier compared to those without state-feedback. It means that the introduction of state-based feedback makes cooperation easier to establish and maintain. Notably, in accordance with the previous conclusion, the local state-feedback can stabilize a full-cooperation state for a sufficiently large $r_0$, as shown in Fig.~\ref{fig2}(l). Specifically, in the absence of punishers (on the CD edge), the public goods game with state-feedback transforms into a coordination game, displaying the bistability of cooperation and defection. Moreover, under the same $r_0$ value, the presence of state-based feedback also enlarges the attractive domain of the full-contribution equilibria, especially the local state-feedback. Consequently, the second-order free-riding problem for the system with state-based feedback is mitigated to some extent, as the stability and attractiveness of punitive behaviors are enhanced.

\begin{figure*}
	\includegraphics[width=0.725\textwidth]{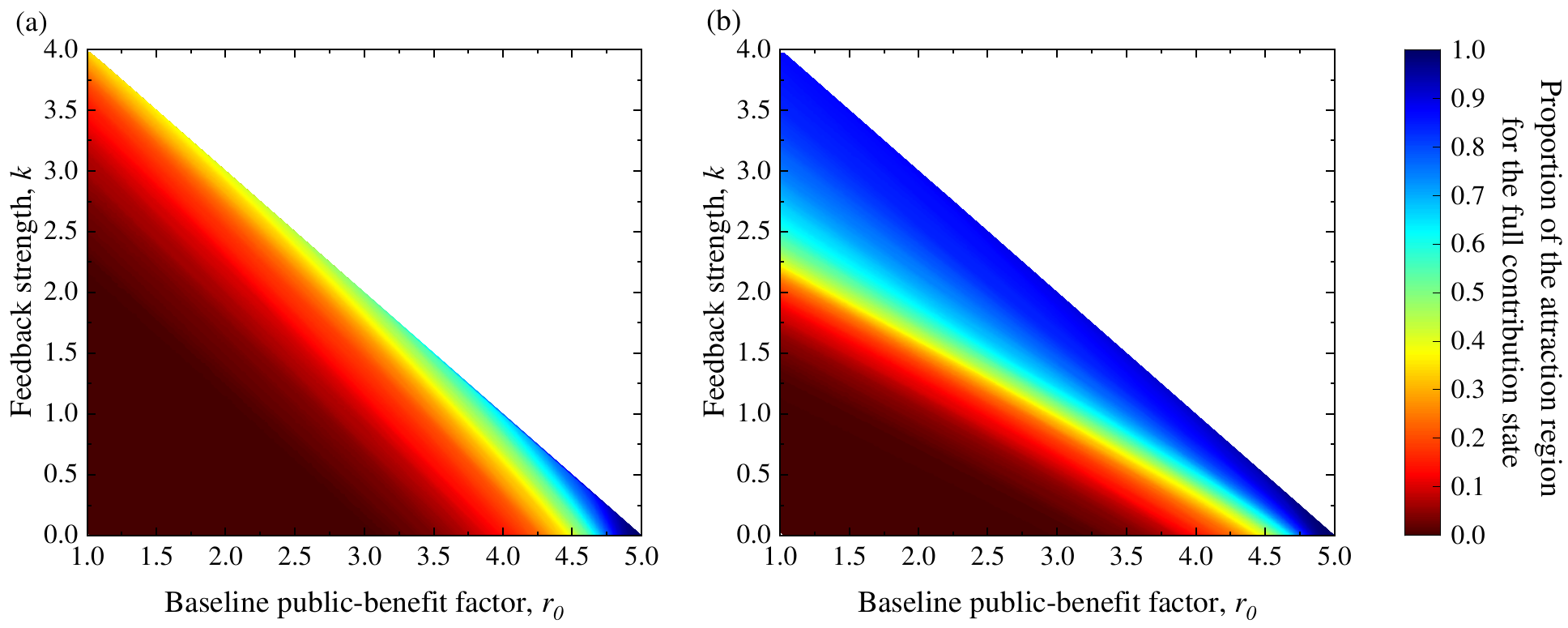}
	\caption{\label{fig4}Effects of $r_0$ and $k$ for the evolution of cooperation under the global state-feedback (left panel) and local state-feedback (right panel). Color maps encode the proportion of attractive regions for the stable full-contribution state in dependence on the baseline multiplication factor $r_0$ and the feedback strength $k$.	Parameters:  $n=5$, $ c=1$, $\gamma=0.01$, and $\beta=0.1$.}
\end{figure*}
Since state-based feedback has a positive consequence, there is no doubt that the stronger the feedback, the more favorable it is for the evolution of cooperation, regardless of the type of feedback.
In Fig.~\ref{fig3}, we substantiate this intuition and explore whether the strength of feedback $k$ influences the  relative advantage of local feedback over global feedback in relieving the cooperation dilemma. We find that independently of the feedback type the system initially evolves to a full-defection state when $k$ is small. However, as $k$ increases,  stable full-contribution states emerge, accompanied by larger basins of attraction. This suggests that a higher value of $k$ increases the likelihood of the system evolving toward a fully contributing state. More significantly, by comparing Fig.~\ref{fig3}(a)-(d) and (e)-(h), local state-feedback demonstrates a superior ability to establish and sustain cooperation compared to global feedback when $k$ is not too small. And, for a sufficiently larger $k$, the system with the local state-feedback can also achieve a full-cooperation state (see Fig.~\ref{fig3}(h)).

Fig.~\ref{fig4} illustrates how the attractive regions of the stable full-contribution state are affected by varying values of $r_0$ and $k$ (which satisfy $1<r_0\le r_0+k<n$) under both global and local state-feedback. In Fig.~\ref{fig4}, blue color indicates the dominance of contributors (where the proportion of attractive domains in the stable full-contribution state is 1), while red color indicates the dominance of defectors (where the proportion of attractive domains in the full-defection state is 1). Other colors represent bistability between full-defectors and full-contributors. For small values of $k$ and $r_0$, the attractive basin of the stable full-contribution state under both global and local state-feedback is nearly missing. As $k$ or $r_0$ increases, the attractive basin of the stable full-contribution state expands for both feedback types. Furthermore, for larger values of $k$ or $r_0$, comparing panels~(a) and (b) of Fig.~\ref{fig4}, we find that feedback based on local interactions encourages cooperation more effectively than relying on global information. Overall, we demonstrate that local state-feedback has general advantages over global feedback in promoting the evolution of cooperation.

\begin{figure*}
	\includegraphics[width=1\textwidth]{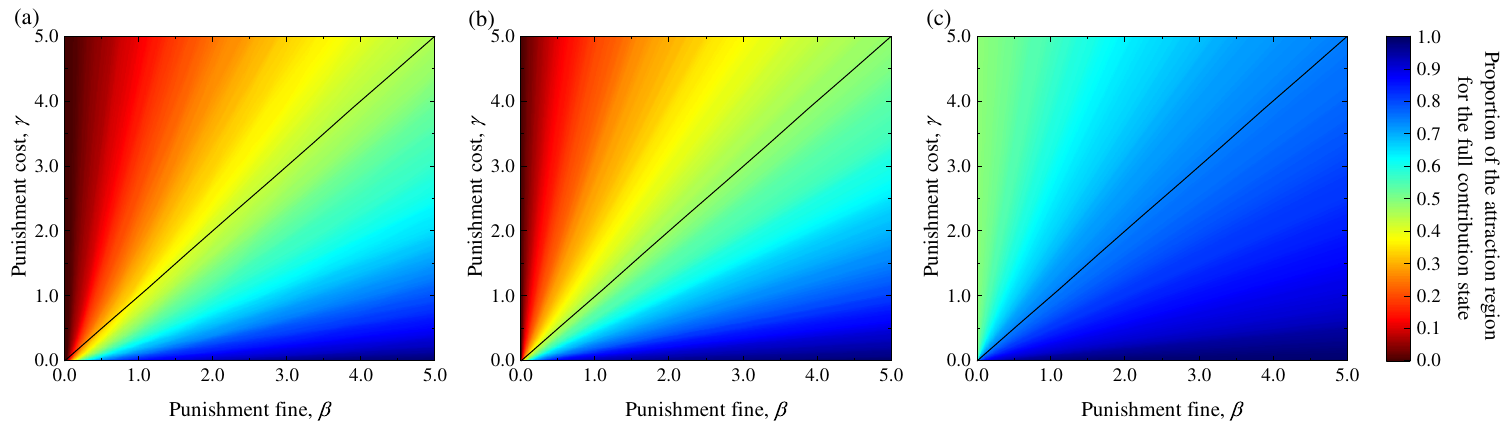}
	\caption{Effects of $\beta$ and $\gamma$ for the evolution of cooperation in the case without feedback (left panel), global state-feedback (middle panel), and local state-feedback (right panel).	Color maps represent the proportion of attractive regions for the stable full-contribution state in dependence on punishment fine $\beta$ and  punishment cost $\gamma$. The diagonal represents the critical threshold at which the cost incurred by the punisher is exactly equal to the loss suffered by the defector due to the punishment (i.e.,  $\gamma=\beta$). Parameters: $n=5$, $ c=1$, $k=0.5$, and $r_0=4.3$.}
	\label{fig5}
\end{figure*}

Next we validate the robustness of the above results to the punishment cost $\gamma$ and the punishment fine $\beta$ in Fig.~\ref{fig5}.
We can observe that the attractive regions of the stable full-contribution state are affected by $\gamma$ and $\beta$ under conditions of no feedback, global feedback, and local feedback, respectively. Evidently, we assume $\gamma<\beta$ in general, as this condition ensures the effectiveness and stability of the punishment mechanism. It is also straightforward that larger $\beta$ and smaller $\gamma$ always facilitate the evolution of cooperation independently of the presence or absence of the feedback mechanism. By comparing the lower regions of the diagonals in Fig.~\ref{fig5}(a)-(c), we observe that the positive effect of state-based feedback, especially the local state-feedback, on cooperative evolution remains unaffected by variations in $\gamma$ and $\beta$. If $\gamma\geq\beta$, the cost of punishment becomes too high for the punisher, eliminating any incentive to enforce the punishment. As a result, the punishment mechanism becomes ineffective, making it more likely for defectors to dominate (see the region above the diagonal in Fig.~\ref{fig5}(a)). Unexpectedly, we find that state-based feedback can partially mitigate the increase in punishers resulting from the failure of the punishment mechanism, particularly in the case of local state-feedback. As illustrated in Fig.~\ref{fig5}(c), when $\gamma\geq\beta$, the attractive region for the stable full-contribution state remains large, covering more than half of the entire phase plane.

\section{\label{sec4}CONCLUSION AND DISCUSSION}

In previous studies on public goods games with punishment, the feedback based on the proportion of contributors to the return of public goods (i.e., the state-based feedback) has often been overlooked. An extreme case of this idea is when the collective income of the group depends sharply on a critical number of cooperators, as it is exemplified in the so-called threshold public good game~\cite{szolnoki_epl10,chen_epl10}.

In this work, to fill this gap, we have investigated how incorporating state-based feedback influences the evolutionary dynamics of strategies in the context of public goods games with peer punishment. Our results have shown that, when state-based feedback is considered, especially local feedback, cooperation can be significantly enhanced. Importantly, when the feedback strength or baseline multiplication factor is sufficiently larger, local state-feedback can enable the system to evolve toward and stabilize full cooperation. Interestingly, our findings have indicated that second-order free-riding can be effectively mitigated under certain conditions, no matter global or local state-feedback is applied. Furthermore, the advantage of state-based feedback remains robust with respect to both the punishment cost and fine.

In this work, we consider the positive feedback effect of collective contributions on the multiplication factor in the public goods game. Accordingly, such state-feedback makes the traditional public goods game nonlinear and this approach has been used in previous studies~\cite{Shi_PhysicaA_2012,Shao_EPL_2019,Lee_ChaosSF_2023,Ma_AMC_2023,Hauert_JTB_2006,Chen_PloSone_2012,Weng_PLA_2021,Hauert_JRSI_2024}. However, different from these works here we not only consider the form of global feedback, but also consider local feedback. Importantly, we focus on the evolutionary dynamics in the state-feedback public goods game with peer punishment, and particularly the synergy effects of such state-feedback and peer punishment. We find that the state-feedback, especially the local version, can strengthen the punishment effects on the evolution of cooperation and alleviate the second-order free rider problem effectively.

Last we note that our study is conducted with an infinite well-mixed population. In reality, populations are usually finite and may exhibit structural complexities. Consequently, in the context of public goods games with punishment in finite and structured populations \cite{Allen_elife_2013,Allen_Nature_2017,WJ_NonlinearDyn_2024,Li_AMC_2024}, exploring the evolutionary dynamics of strategies under the suggested state-based feedback can offer an important and challenging avenue for future research. On the other hand, various incentive strategies have been also explored in the public goods game to investigate the evolution of cooperation, such as social exclusion \cite{Sasaki_PRSB_2013} and reward \cite{Santos_PRSB_2015,Sasaki_BL_2014}. It would be promising to explore the potential synergistic effects of these strategies in combination with state-based feedback in future studies.

\begin{acknowledgments}
	This research was supported by the National Natural Science Foundation of China (Grants No. 62473081 and No. 62036002), the Sichuan Science and Technology Program (Grant No. 2024NSFSC0436), and by the National Research, Development and Innovation Office (NKFIH) under Grant No. K142948.
\end{acknowledgments}

\section*{DATA AVAILABILITY STATEMENT}
No new data were created or analyzed in this study.

\appendix

\section{\label{suppA}Evolutionary dynamics without state-based feedback}
In the classic public goods game with punishment, we assume that the synergy of the public pool on individual contributions is constant~\cite{Sasaki_PRSB_2013,Hauert_DCDSSB_2004}. Specifically, let $r = r_0$,  representing a baseline multiplication factor that does not vary with the level of contributions. The replicator equations are
\begin{equation}\label{sys_0}
	\left\{
	\begin{aligned}
		\dot{x}&=x(1-x-z)\left[z (\beta+\gamma)(n-1)+\frac{r_{0}}{n}c -c \right], \\
		\dot{z}&=z(1-x-z)\left[ z(\beta+\gamma)(n-1)+\frac{r_{0}}{n}c -c -\gamma(n-1)\right].
	\end{aligned}
	\right.
\end{equation}
We know that $(0,0)$, $(0,\psi)$, and $(x',z')$ are equilibrium points, where $\psi=\frac{\gamma n(n-1)+c(n-r_{0})}{(\beta+\gamma)n(n-1)}$ and $x'+z'=1$. The Jacobian matrices at each possible equilibrium are
\begin{equation}\nonumber
	J(0,0)=
	\left[ \begin{array}{cc}
		\frac{r_{0}}{n}c-c&0\\
		\noalign{\medskip}0&\frac {r_{0}}{n}c-c-\gamma \left( n-1\right)
	\end{array} \right],
\end{equation}
\begin{equation}\nonumber
	\begin{split}
		J\left(0,\psi\right)
		=
		\left[
		\begin{array}{cc}
			{ \gamma\frac{
					\beta{n}(n-1)-c(n-r_{0}) }{(\beta+\gamma)n}}&0\\
			\noalign{\medskip}0&
			\frac{ (\beta {n}(n-1)-c(n-r_{0}))(\gamma n(n-1)+c(n-r_{0})) }{(\beta+\gamma)
				n^{2}(n-1)}
		\end{array} \right],
	\end{split}
\end{equation}
and
\begin{widetext}
\begin{equation}\nonumber
		J(x',z')
		=
		\left[ \begin{array}{cc}
			- \left( 1-z' \right)  \left( z' \left( \beta
			+\gamma \right)  \left( n-1 \right) + {\frac{r_{0} }{n}}c-c \right) &- \left( 1-z' \right)  \left( z' \left( \beta
			+\gamma \right)  \left( n-1 \right) + {\frac{r_{0}}{n}}c-c \right) \\
			\noalign{\medskip}
			-z' \left( z' \left(
			\beta+\gamma \right)  \left( n-1 \right) +{\frac{r_{0} }{n}}c-c-\gamma \left( n-1 \right)  \right) &-z' \left( z
			\left( \beta+\gamma \right)  \left( n-1 \right) +{\frac{r_{0} }{n}}c-c-\gamma\left( n-1 \right)  \right)
		\end{array} \right].
\end{equation}
\end{widetext}

Next, we analyze the local stability of the equilibria. For convenience, we set $z^{*}_{N}=\frac{ n-r_{0}}{\beta n(n-1)}c$. Since $1<r<n$, thus $z^{*}_{N}>0$.
\begin{enumerate}
	\item
	When \( 0 < z^{*}_{N} < 1 \), i.e, $\beta>\frac{n-r_{0}}{n(n-1)}c>0$, the equilibria $(0,0)$, $(0,\psi)$, and $(x',z')$ exist.
	
	\quad The eigenvalues of the Jacobian matrix at $(0,0)$ are $\frac{r_{0}}{n}c - c - \gamma(n-1)$ and $\frac{r_{0}}{n}c - c$, both of which are negative, indicating that \((0,0)\) is a stable node.
	
	For the equilibrium \((0,\psi)\), the eigenvalues of the Jacobian matrix are \(\gamma\frac{\beta n(n-1) - c(n - r_{0})}{(\beta + \gamma)n}\) and \(\frac{(\beta n(n-1) - c(n - r_{0}))(\gamma n(n-1) + c(n - r_{0}))}{(\beta + \gamma)n^{2}(n-1)}\). Both of these eigenvalues are positive, making \((0,\psi)\) an unstable node.
	
	\quad The eigenvalues of the Jacobian matrix at \((x', z')\) are \(0\) and \(-z'\beta(n-1) + \frac{n-r_{0}}{n}c\). Consequently, the stability of $(x',z')$ is examined by using a straightforward perturbation analysis \cite{Sasaki_PRSB_2013,Huang_JTB_2018}. For any point $(x,z)$, where $x+z<1$ and $(x,z)$ is closed to $(x',z')$, $\frac{d(x+z)}{dt}=(x+z)(1-x-z)(z(\beta+\gamma)(n-1)+\frac{r_0}{n}c-c+\frac{z}{x+z}\gamma(1-n))$.
	\begin{enumerate}
		\item When \(0 \leq z' < z^{*}_{N}\), $\frac{d(x+z)}{dt}$ is negative iff	$f_N(x,z)\triangleq z(\beta+\gamma)(n-1)+\frac{r_0}{n}c-c+\frac{z}{x+z}\gamma(1-n)<0$. Thus, under small perturbations at $(x',z')$, $(x,z)$  will always evolve away from the state $(x',z')$.
		\item  When \(z^{*}_{N} < z' \leq 1\), $\frac{d(x+z)}{dt}$ is positive iff	$f_N(x,z)>0$. In this case, under small perturbations at $(x',z')$, $(x,z)$ will always evolve into the state $(x',z')$.
	\end{enumerate}
	Therefore, for a fixed $z^*_{N}$ (where $0<z^*_{N}<1$), the equilibrium $(x',z')$ is unstable when $0 \leq z' < z^{*}_{N}$ and locally stable when $z^{*}_{N} < z' \leq 1$.
	
	\item When $z^{*}_{N}>1$, i.e, $\beta<\frac{n-r_{0}}{n(n-1)}c$, the equilibria $(0,0)$ and $(x',z')$ exist.
	
	\quad The point $(0,0)$ is a stable node. However, since $-z'\beta(n-1)-\frac{r_{0}}{n}c+c$ is always positive, the equilibrium $(x',z')$  is unstable.
\end{enumerate}

\section{\label{suppB}Evolutionary dynamics with global state-feedback}

We then introduce the global feedback of contributors in the population on the multiplication factor into the classical public goods game with punishment. We first investigate how the multiplication factor $r$ increases with the number of contributors in the entire population. Specifically, we assume that $r=r_0+k(x+z)$, where $k>0$ represents the feedback strength. According to Eq.~\eqref{average payoff1}, the expected payoffs for cooperators $C$, defectors $D$, and punishers $P$ are obtained:
\begin{equation}\nonumber
		P_{C}
		=
		\frac{ r_{0}+k(x+z) }{n}c(x+z)(n-1)+\frac{ r_{0}+k(x+z) }{n}c-c,
\end{equation}
\begin{equation}\nonumber
		P_{D}
		=
		\frac{ r_{0}+k(x+z) }{n}c(x+z)(n-1)-\beta(n-1)z,
\end{equation}
and
\begin{equation}\nonumber
	\begin{split}
		P_{P}=&
		\frac{r_{0}+k(x+z)} {n} c(x+z)(n-1)  \\
		& +\frac{ r_{0}+k(x+z) }{n} c-c -\gamma(n-1)y,
	\end{split}
\end{equation}respectively.
Then, the replicator dynamics are described by the differential equation system:
\begin{equation}\nonumber
	\left\{
	\begin{aligned}
		\dot{x}=&x(1-x-z)\left[z (\beta+\gamma)(n-1)+\frac{r_{0}+k(x+z)}{n}c -c \right], \\
		\dot{z}=&z(1-x-z)\cdot\left[ z(\beta+\gamma)(n-1)+\frac{r_{0}+k(x+z)}{n}c \right.  \\  & \left.-c -\gamma(n-1)\vphantom{\sum_{a}}\right].
	\end{aligned}
	\right.
\end{equation}

There are at most two isolated equilibria within a local area, namely $(0,0)$ and $\left(0, \phi\right)$, where $\phi=\frac{(n-r_{0})c+\gamma n(n-1)}{(\beta+\gamma)n(n-1)+kc}$. Besides, each point $(x',z')$ that satisfies $x'+z'=1$ ($1\ge x',z'\ge0$) is an equilibrium.
The Jacobian matrices at each possible equilibrium  are
\begin{equation}\nonumber
	J(0,0)=
	\left[ \begin{array}{cc}
		\frac{r_{0}}{n}c-c&0\\
		\noalign{\medskip}0&\frac {r_{0}}{n}c-c-\gamma \left( n-1\right)
	\end{array} \right],
\end{equation}
\begin{widetext}
\begin{equation}\nonumber
	J\left(0,\phi\right)
	=
	\left[
	\begin{array}{cc}
		{\frac{ \gamma\left( n-1 \right)
				\left( \beta{n}(n-1)-c(n-r_{0}-k)\right) }{(\beta+\gamma)n(n-1)+kc}}&0\\
		\noalign{\medskip}{
			\frac { kc\left( \beta {n}(n-1)-c(n-r_{0}-k) \right)
				\left( \gamma\,{n}(n-1)+c(n-r_{0})\right) }{ n\left(
				(\beta+\gamma){n}(n-1)+kc\right)^{2}}}&
		\frac{ (\beta {n}(n-1)-c(n-r_{0}-k))(\gamma n(n-1)+c(n-r_{0})) }{n ( (\beta+\gamma)
			{n}(n-1)+kc)}
	\end{array} \right],
\end{equation}
\end{widetext}
and
\begin{equation}\nonumber	
	\begin{split}
		J(x',z')
		=
		\left[ \begin{array}{cc}
			J_1 &J_1 \\
			\noalign{\medskip}
			J_2&J_2
		\end{array} \right],
	\end{split}
\end{equation}
where
$$
J_1=- \left( 1-z' \right)  \left( z' \left( \beta
+\gamma \right)  \left( n-1 \right) + {\frac{r_{0}+k  }{n}}c-c \right)
$$
and
$$
J_2=-z' \left( z' \left( \beta+\gamma \right)  \left( n-1 \right) +{\frac{ r_{0}+k}{n}}c-c-\gamma\, \left( n-1 \right)  \right).
$$
Next, we analyze the local stability of the equilibria. For convenience, we define $z^{*}_{G}=\frac{n-r_{0}-k}{\beta n(n-1)}c$, which is clearly positive.

\begin{enumerate}
	\item
	When $0<z^{*}_{G}<1$, i.e., $n-r_{0}-\frac{\beta n(n-1)}{c}<k<n-r_{0}$, the equilibria $(0,0)$, $\left(0, \phi\right)$, and  $(x',z')$ exist.
	\quad For the equilibrium $(0,0)$, the eigenvalues of the Jacobian are $\frac {r_{0}}{n}c-c<0$ and $\frac {r_{0}}{n}c-c-\gamma \left( n-1\right)<0$. Note that the value of $z^{*}_{G}$ does not affect the existence of $(0,0)$ or the sign of its eigenvalues,  ensuring that  $(0,0)$ is always a stable node.
	
	\quad Consider the equilibrium  $\left(0, \phi\right)$. Given that $\beta{n}(n-1)-c(n-r_{0}-k)>0$, the eigenvalues of the Jacobian at this point are both positive, i.e.,
	$$
	\frac{ \gamma\left( n-1 \right)
		\left( \beta{n}(n-1)-c(n-r_{0}-k)\right) }{(\beta+\gamma)n(n-1)+kc} >0
	$$
	and
	$$
	\frac{ (\beta {n}(n-1)-c(n-r_{0}-k))(\gamma n(n-1)+c(n-r_{0})) }{n ( (\beta+\gamma)
		{n}(n-1)+kc)}>0.
	$$
	Therefore, $\left(0,\phi\right) $ is an unstable node.
	
	\quad For the equilibrium $(x',z')$, the eigenvalues of the Jacobian are
	\begin{equation}\nonumber
		0\quad {\rm and}\quad
		-z'\beta (n-1)+\frac{n-r_{0}-k}{n}c\,.
	\end{equation}
	For any point $(x,z)$, where $x+z<1$ and $(x,z)$ is closed to $(x',z')$, $\frac{d(x+z)}{dt}=(x+z)(1-x-z)(z(\beta+\gamma)(n-1)+\frac{r_0+k(x+y)}{n}c-c+\frac{z}{x+z}\gamma(1-n))$.
	\begin{enumerate}
		\item When $0 \leq z' < z^{*}_{G}$, $\frac{d(x+z)}{dt}$ is negative iff
		$f_G(x,z)\triangleq z(\beta+\gamma)(n-1)+\frac{r_0+k(x+z)}{n}c-c+\frac{z}{x+z}\gamma(1-n)<0$. Thus, under small perturbations at $(x',z')$, $(x,z)$  will always evolve away from the state $(x',z')$.
		\item  When \(z^{*}_{G} < z' \leq 1\), $\frac{d(x+z)}{dt}$ is positive iff
		$f_G(x,z) >0$. In this case, under small perturbations at $(x',z')$, $(x,z)$ will always evolve into the state $(x',z')$.
	\end{enumerate}
	Therefore, for a fixed $z^{*}_{G}$ (where $0<z^{*}_{G}<1$), the equilibrium $(x',z')$ is unstable when \(0 \leq z' < z^{*}_{G}\) and locally stable when \(z^{*}_{G} < z' \leq 1\).
	
	\item
	When $z^{*}_{G}>1$, namely, $k<n-r_{0}-\frac{\beta n(n-1)}{c}$, the equilibria $(0,0)$ and  $(x',z')$ exist.
	
	\quad As before, $(0,0)$ is a stable node.
	The eigenvalues of the Jacobian matrix at $(x',z')$ are
	$$
	0\quad {\rm and} \quad -z'\beta (n-1)+\frac{n-r_{0}-k}{n}c>0.
	$$
	Therefore, the equilibrium $(x',z')$, where $x'+z'=1$, is unstable.
\end{enumerate}

\section{\label{suppC}Evolutionary dynamics with local state-feedback}

For local state-feedback, the multiplication factor $r$ is positively correlated with the frequency of contributors participating in the public goods game, expressed as
\begin{equation}
	r=r_{0}+k\frac{s'}{n},
\end{equation}where $s'$ denotes the number of contributors among $n$ participants in the public goods game.
The expect payoffs for cooperators $C$, defectors $D$, and punishers $P$ are obtained:
\begin{equation}\nonumber
	\begin{split}
	P_{C}
		=&
		\frac{k(n-1)c}{n^2}((n-2)y-2n+1)y\\
		&-\frac{r_0(n-1)c}{n}y+(k+r_0-1)c, \\
	P_{D}
		=&
		\frac{(n-1)c}{n}(x+z)((x+z)k+r_0)\\
		&+\frac{(n-1)k}{n^2}(2y-1)-\beta(n-1)z,
	\end{split}
\end{equation}
and
\begin{equation}\nonumber
	\begin{split}
		P_{P}
		=&
		\frac{k(n-1)c}{n^2}((n-2)y-2n+1)y\\
		&-\frac{r_0(n-1)c}{n}y+(k+r_0-1)c -\gamma(n-1)y.
	\end{split}
\end{equation}
Accordingly, the replicator equation system can be written as
\begin{equation}\nonumber
	\left\{
	\begin{aligned}
		\dot{x}=&x(1-x-z)\left[z (\beta+\gamma)(n-1)+\frac{2ck}{n^{2}}(x+z)\left(n -1\right)\right.  \\
		& \left.+\frac{r_{0}}{n}c+\frac{k}{n^{2}}c-c \right], \\
		\dot{z}=&z(1-x-z)
		\left[z (\beta+\gamma)(n-1)+\frac{2ck}{n^{2}}(x+z)\left(n -1\right)\right.  \\
		& \left.+\frac{r_{0}}{n}c+ \frac{k}{n^{2}}c-c -\gamma(n-1)\right].
	\end{aligned}
	\right.
\end{equation}

There are at most three isolated equilibria within a local area: $(0,0)$, $\left(0, \varsigma\right) $, and $\left(\frac{n^{2}- nr_{0}-k}{2k \left( n-1 \right)},0 \right) $, where $\varsigma=\frac{\gamma n^{2}\left( n-1 \right) +\left( n^{2}- nr_{0}-k\right)c}{(\beta+\gamma)n^{2}(n-1) +2kc\left(n-1\right)}$.
Besides, each point $(x',z')$ that satisfies $x'+z'=1$ ($0\le x',z'\le1$) is an equilibrium.
The Jacobian matrices of the system at these fixed points and their corresponding eigenvalues are
\begin{equation}\nonumber
	J(0,0)=
	\left[ \begin{array}{cc}
		-\frac{n^{2}-nr_{0}-k }{n^{2}}c&0\\
		0&-\frac{n^{2}-nr_{0}-k }{n^{2}}c-\gamma (n-1)
	\end{array} \right],
\end{equation}
\begin{equation}\nonumber
	J\left(0,\varsigma\right) \\
	=
	\left[ \begin{array}{cc}
		\gamma\frac{\beta n^{2}(n-1)+\left( 2k(n-1)-(n^{2}-nr_{0}-k)\right)c}{(\beta+\gamma)n^{2}+2kc}&0\\
		\frac{2kc }{ (\beta+\gamma)n^{2}+2kc }J_3&J_3
		\end {array} \right],
\end{equation}
\begin{widetext}
	\begin{equation}\nonumber
		\begin{split}
			J\left(\frac{n^{2}- nr_{0}-k}{2k \left( n-1 \right)},0\right)
			=
			\left[ \begin {array}{cc}
			\frac{c\left(n^{2}- nr_{0}-k\right)\left( 2k(n-1)-(n^{2}-nr_{0}-k)\right)}{2kn^{2}\left(n-1\right)}&\frac{((\beta+\gamma)n^{2}+2kc)\left(n^{2}- nr_{0}-k\right)\left( 2k(n-1)-(n^{2}-nr_{0}-k)\right)}{4k^{2}n^2\left( n-1
				\right) }\\
			0&-\frac {\gamma\left( 2k(n-1)-(n^{2}-nr_{0}-k)\right) }{2k}
			\end {array} \right],
		\end{split}
	\end{equation}
\end{widetext}
and
\begin{equation}\nonumber
	J(x',z')=
	\left[ \begin{array}{cc}
		J_4 &J_4 \\
		J_5 &J_5
	\end{array} \right],
\end{equation}
where
\begin{equation}\nonumber
	\begin{split}
		J_3=& \left[ \beta n^{2}(n-1)+\left( 2k(n-1)-(n^{2}-nr_{0}-k)\right)c\right] \\
		&\cdot\frac{\left(\gamma n^{2}(n-1)+c(n^{2}-nr_{0}-k)\right)}{n^{2}\left(n-1\right) \left( (\beta+\gamma)n^{2}+2kc\right)},\\
		J_4=&- \left(1-z\right)\left[ \vphantom{\sum_{a}} z\left(\beta+\gamma\right)\left( n-1\right) \right.  \\
		& \left.+ {\frac{2k(n-1)-(n^{2}-r_{0} n-k)}{n^{2}}}c\right]
	\end{split}
\end{equation}
and
\begin{equation}\nonumber
	\begin{split}
		J_5=&-z \left[ \vphantom{\sum_{a}} z\left(\beta+\gamma\right)\left( n-1\right) \right.  \\
		& \left.+{\frac{2k(n-1)-(n^{2}-r_{0} n-k)}{n^{2}}}c-\gamma(n-1)\right] .
	\end{split}
\end{equation}

Next, we analyze the local stability of the equilibria. For convenience, we define $z^{*}_{L}=\frac{(n^{2}-nr_{0}-k)-2k(n-1)}{\beta n^{2}(n-1)}c$, which is greater than $-\frac{2k}{\beta n^{2}}c$.
\begin{enumerate}
	\item
	When $0<z^{*}_{L}<1$, i.e., $\frac{n(n-r_{0})}{2n-1}-\frac{\beta n^{2}(n-1)}{c(2n-1)}<k<\frac{n(n-r_{0})}{2n-1}$,
	the equilibria $(0,0)$, $\left(0, \varsigma\right)$, and $(x',z')$ exist.
	
	\quad For the equilibrium $(0,0)$, the eigenvalues of the Jacobian are $-\frac{n^{2}-nr_{0}-k }{n^{2}}c<0$ and $-\frac{n^{2}-nr_{0}-k }{n^{2}}c-\gamma (n-1)<0$, ensuring that  $(0,0)$ is a stable node.
	
	\quad At the equilibrium  $\left(0, \varsigma\right)$, the eigenvalues of the Jacobian matrix are both positive:
	$$
	\gamma\cdot\frac{\beta n^{2}(n-1)+\left( 2k(n-1)-(n^{2}-nr_{0}-k)\right)c}{(\beta+\gamma)n^{2}+2kc}>0
	$$
	and
	$$
	J_3>0.
	$$
	Therefore, $\left(0, \varsigma\right)$ is an unstable node.
	
	\quad For the equilibria $(x',z')$, where $x'+z'=1$, the eigenvalues of the Jacobian matrix are
	\begin{equation}\nonumber
		0\quad {\rm and}\quad
		-z'\beta(n-1)- {\frac{2k(n-1)-(n^{2}-r_{0} n-k)}{n^{2}}}c.
	\end{equation}
	For any point $(x,z)$, where $x+z<1$ and $(x,z)$ is closed to $(x',z')$, $\frac{d(x+z)}{dt}=(x+z)(1-x-z)(z (\beta+\gamma)(n-1)+\frac{2kc}{n^{2}}\left(n -1\right)(x+z)+\frac{r_{0}}{n}c+ \frac{k}{n^{2}}c-c +\frac{z}{x+z}\gamma(1-n))$.
	\begin{enumerate}
		\item When $0 \leq z' < z^{*}_{L}$, $\frac{d(x+z)}{dt}$ is negative iff $f_L(x,z)\triangleq z (\beta+\gamma)(n-1)+\frac{2kc}{n^{2}}\left(n -1\right)(x+z)+\frac{r_{0}}{n}c+ \frac{k}{n^{2}}c-c +\frac{z}{x+z}\gamma(1-n)<0$. Thus, under small perturbations at $(x',z')$, $(x,z)$ will always evolve away from the state $(x',z')$.
		\item  When \(z^{*}_{L} < z' \leq 1\), $\frac{d(x+z)}{dt}$ is positive iff	$f_L(x,z) >0$. In this case, under small perturbations at $(x',z')$, $(x,z)$ will always evolve into the state $(x',z')$.
	\end{enumerate}
	Therefore, for a fixed $z^{*}_{L}$ (where $0<z^{*}_{L}<1$), the equilibrium $(x',z')$ is unstable when \(0 \leq z' < z^{*}_{L}\) and stable when \(z^{*}_{L} < z' \leq 1\).
	
	\item
	When $z^{*}_{L}>1$, i.e., $k<\frac{n(n-r_{0})}{2n-1}-\frac{\beta n^{2}(n-1)}{c(2n-1)}$.
	In this case, only the equilibria $(0,0)$ and $(x',z')$ exist.
	
	\quad For the equilibrium $(0,0)$, the eigenvalues of the Jacobian matrix are negative, confirming that $(0,0)$ is a stable node.
	
	\quad However, for the equilibrium $(x',z')$, where $x'+z'=1$, both eigenvalues of the Jacobian are non-negative, indicating that $(x',z')$ is unstable.
	
	\item
	When $-\frac{2k}{\beta n^2}c<z^{*}_{L}<0$, i.e., $\frac{n(n-r_{0})}{2n-1}<k<n-r_{0}$,
	the equilibria $(0,0)$, $\left(0, \varsigma\right) $, $\left(\frac{n^{2}- nr_{0}-k}{2k \left( n-1 \right)},0 \right) $, and $(x',z')$ exist.
	
	\quad For the equilibrium $(0,0)$, the eigenvalues of the Jacobian matrix are negative, making $(0,0)$ a stable node.
	The equilibrium  $\left(0, \varsigma\right) $ has two positive eigenvalues, so it is an unstable node.
	
	\quad For the equilibrium $\left(\frac{n^{2}- nr_{0}-k}{2k \left( n-1 \right)},0 \right)$, the Jacobian matrix has one positive and one negative eigenvalue, i.e.,
	$$
	\frac{c\left(n^{2}- nr_{0}-k\right)\left( 2k(n-1)-(n^{2}-nr_{0}-k)\right)}{2kn^{2}\left(n-1\right)}>0
	$$
	and
	$$
	-\frac {\gamma\left( 2k(n-1)-(n^{2}-nr_{0}-k)\right) }{2k}<0.
	$$
	Therefore, $\left(\frac{n^{2}- nr_{0}-k}{2k \left( n-1 \right)},0 \right)$ is an unstable saddle point.
	
	\quad For the equilibrium $(x',z')$, where $x'+z'=1$,  the Jacobian matrix has one zero eigenvalue and one negative eigenvalue.
	For any point $(x,z)$, where $x+z<1$ and $(x,z)$ is closed to $(x',z')$, $\frac{d(x+z)}{dt}=(x+z)(1-x-z)(z (\beta+\gamma)(n-1)+\frac{2kc}{n^{2}}\left(n -1\right)(x+z)+\frac{r_{0}}{n}c+ \frac{k}{n^{2}}c-c +\frac{z}{x+z}\gamma(1-n))$.
	Obvious, $\frac{d(x+z)}{dt}$ is positive iff $f_L(x,z)>0$. Thus, under small perturbations at $(x',z')$, $(x,z)$ will always evolve into the state $(x',z')$. Therefore, the equilibrium $(x',z')$ is locally stable.
\end{enumerate}

\end{document}